\documentclass[12pt,twoside,leqno]{article}
\usepackage{amsfonts}
\usepackage{epsfig}
\usepackage{amssymb,amsmath,bm}
\usepackage{subfigure}
\usepackage{amsthm}
\usepackage{mathrsfs}
\usepackage{color} 
\usepackage[section]{algorithm}
\usepackage{fancyhdr}
\theoremstyle{plain}
\newtheorem{proposition}{Proposition}

\newcounter{hypA}

\begin{document}

\bigskip

\begin{center}

{\Large \textbf{Sequential Monte Carlo Methods for Option Pricing}}

\bigskip

BY AJAY JASRA and PIERRE DEL MORAL \\[0pt]

\emph{Department of Mathematics, Imperial College London, SW7
2AZ, London, UK}, \texttt{Ajay.Jasra@ic.ac.uk}%
\\[0pt]

\emph{Centre INRIA Bordeaux et Sud-Ouest \& Institut de Math\'ematiques de Bordeaux, Universit\'e de Bordeaux I, 33405, France,},\\ \texttt{Pierre.Del-Moral@inria.fr}%
\\[0pt]

\end{center}

\begin{abstract}
In the following paper we provide a review and development of sequential
Monte Carlo (SMC) methods for option pricing. SMC are a class of Monte Carlo-based algorithms, that are designed to approximate expectations w.r.t a sequence of related probability measures. These approaches
have been used, successfully, for a wide class of applications in engineering,
statistics, physics and operations research.
SMC methods are highly suited to many option pricing problems and sensitivity/Greek
calculations due to the nature of the sequential simulation. However, it
is seldom the case that such ideas are explicitly used in the option pricing
literature. This article provides an up-to date review of SMC methods, which
are appropriate for option pricing. In addition, it is illustrated how a number of existing
approaches for option pricing can be enhanced via SMC. Specifically,
when pricing the arithmetic Asian option w.r.t a complex stochastic volatility
model, it is shown that SMC methods provide additional strategies to improve estimation.
\textbf{Short Title:} SMC for Option Pricing\
\end{abstract}


\thispagestyle{plain}

\section{Introduction}

Monte Carlo methods have been extensively used in option pricing since the paper of \cite{boyle}. Subsequently, there have been a wide variety
of Monte Carlo approaches applied: quasi Monte Carlo
(QMC), stratification, importance sampling (IS), control and antithetic variates, etc; see \cite{glasserman} for a thorough introduction.

The importance of Monte Carlo for option pricing, against other
numerical approaches, is the ability to deal with high-dimensional integrals.
This is either in the time parameter of the derivative (path dependent options), or in the dimension of the underlying. The rate of convergence is $O(1/\sqrt{N})$, $N$ being the number of simulated samples,
supposedly, independent of the dimension. In addition, the methods
are straight-forwardly extended to:
\begin{itemize}
\item{complex stochastic volatility (SV) models (e.g.~\cite{bns})}
\item{complicated financial derivatives (e.g.~\cite{broadie}).}
\end{itemize}
SV models are particularly useful to realistically replicate price
dynamics. The latter point is relevant due to an increase in the
volume traded of these instruments. Monte Carlo may also be used to
calculate sensitivities/Greeks (e.g.~\cite{fournie}).
As frequently noted in the option pricing
literature, standard Monte Carlo estimates can suffer from a high level of
variability, but can
be improved using some of the methods mentioned in the above paragraph.

We have thus stated that Monte Carlo methods are an important tool for option
pricing,
but can suffer from high variability. An often used technique
to deal with this problem is IS (e.g.~\cite{glasserman1}).
As is well known, the idea is to change the dominating measure such that the
resulting Monte Carlo estimation benefits from a lower variance. In many
financial applications the simulation is sequential, that is, the underlying
is sampled at discrete time-points. This often yields Radon-Nikodym derivatives
that can be re-calculated sequentially in time. It is also well-known 
(e.g.~\cite{doucet2,glasserman,liu}) that the variance of these weights increases with time; hence IS needs to be improved in order
to yield accurate estimates of the option-price.

This above problem can typically be solved using sequential Monte
Carlo methods. This is a class of IS methods that are extensively
used in engineering, statistics and physics. However, to our knowledge, these
ideas are seldom used in the option pricing literature (see \cite{defreitas,florescu,johannes,rambh,zhao} for the few applications we were able to find). The purpose of this
article is thus two-fold:
\begin{enumerate}
\item{To provide an up-to date literature review of SMC methods, particularly
focussed on option pricing and sensitivity analysis.\label{enumi:1}}
\item{To illustrate that such methods can help to push the boundaries of
the models for which option prices can be calculated, accurately.\label{enumi:2}}
\end{enumerate}
In terms of \ref{enumi:1}, it seems that such methods are not well understood
in the financial engineering literature, or at least the benefits of their
application in option pricing is not appreciated. Therefore, we aim to review such
methods and illustrate their use as well as improvement over standard approaches.
Indeed, in some cases it is even possible to calculate option prices that
are not subject to time discretization error \cite{fearnhead}; this
requires SMC methods.
In relation to \ref{enumi:2}, an SMC algorithm is introduced
to compute the value of arithmetic Asians, when the underlying is modelled
by a stochastic volatility model. The volatility follows a non-Gaussian
Ornstein-Uhlenbeck process \cite{bns}.
This problem cannot be easily solved using ordinary
IS and deterministic methods; see Section \ref{sec:pricingasians} for further details. 
Note, that it should not be
seen that SMC methods are competitors to existing methods in option pricing,
but simply that they enrich the methodology that can be used: in many cases SMC can be combined with existing ideas such as stratification (see e.g.~\cite{etore} for some recent work).

It is remarked that the application of SMC methods can substantially reduce
the variance of ordinary Monte Carlo and IS estimators; this at the cost
of an increase the computational cost. In other words, the methods can be
the most accurate in comparison to other approaches.
Due to the above statements, the methods reviewed here in most cases could
not be used at high frequency, but if solutions are required in minutes
can be actively used in finance (although see \cite{lee} for computational
hardware that may make the methods even more applicable). Note, also the ideas differ from parametric IS, where the
proposal lies in a parametric class, and is found to minimize some criterion
(e.g.~as in the cross entropy method see \cite{rubinstein} for details). Since these methods are not SMC techniques, we do not review them
here.

This article is structured as follows. In Section \ref{sec:motivatingex}
we discuss an example to motivate the application of SMC methods for
option pricing. In Section \ref{sec:smcmethods}, SMC methods are detailed along with some of the
latest developments. Illustrations are given on various examples to show
that SMC methods can enhance existing Monte Carlo approaches in option pricing. In Section \ref{sec:pricingasians} an original SMC approach is designed for pricing Asian options, using the Barndorff-Nielsen \& Shephard (BNS) SV model \cite{bns}. In Section \ref{sec:summary} the article is concluded and some avenues for future research are discussed. There is an appendix which gives the proof
of a result (in Section \ref{sec:smcmethods}) and some details from the example in Section
\ref{sec:pricingasians}.

\section{Motivating Example}\label{sec:motivatingex}

\subsection{Some Conventions}

Some conventions are given. Recall that a Monte Carlo procedure
simulates $N$ independent samples, $X^{(i)}$, from a density, $\pi$ and estimates $\pi-$integrable functions $h$ via
$$
\frac{1}{N}\sum_{i=1}^N h(X^{(i)}).
$$
Importance sampling follows much the same formula, except sampling from an
$q \gg \pi$ and using the estimate
$$
\sum_{i=1}^N w^{(i)}h(X^{(i)})
$$
with $w^{(i)}=d\pi/dq(X^{(i)})$.
The following notation is used. For any $(i,j)\in \mathbb{Z}^+$, $i\leq j$,
$x_{i:j}:=(x_i,\dots,x_j)$. A process is
written $\{X_t\}_{t\in[0,T]}$. A measurable space is denoted $(E,\mathcal{E})$.
Given a sequence of spaces $E_0,\dots,E_n$ (resp.~$\sigma-$algebras $\mathcal{E}_0,\dots,\mathcal{E}_n$) the product space is written as $E_{[0,n]}$ (resp.~product $\sigma-$ algebra
$\mathcal{E}_{[0,n]}$).
For a probability $\pi$ and $\pi-$integrable function $h$, the notation
$\pi(h):=\int h(x)\pi(x)dx$ is sometimes used. The dirac measure on $\{x\}$ is written $\delta_x(dx')$. Probability densities are often assigned a standard
notation $p$.
Expectations are written generically as $\mathbb{E}$
and a subscript is added, if it is required to denote dependence upon a measure/point.
Also, for $p\in\mathbb{Z}^+$, $\mathbb{T}_p=\{1,\dots,p\}$. 

\subsection{Price Process}
Throughout, the price process $\{S_t\}_{t\in [0,T]}$ $S_t\in\mathbb{R}^d$
follows a general jump-diffusion of the form
$$
dS_t = \mu(S_t)dt +  V_t\sigma(S_t) dZ_t
$$
where $\{Z_t\}_{t\in[0,T]}$ is a L\'evy process, and the volatility $\{V_t\}_{t\in[0,T]}$,
may be deterministic, or may follow
$$
dV_t = \alpha(V_t)dt + \beta(V_t) dU_t
$$
where $\{U_t\}_{t\in[0,T]}$ is a L\'evy process, that may be correlated with
$\{Z_t\}_{t\in[0,T]}$. This permits a wide-class of stochastic volatility
models, which can accurately replicate the stylized features of returns data
(e.g.~\cite{bns}).
All expectations are taken w.r.t an equivalent, structure preserving,
Martingale measure, $\mathbb{Q}$; this will exist for our examples.

\subsection{Barrier Options}\label{sec:barrieroptions}
The first example is the calculation of barrier options. These
are derivatives for which the payoff may be zero, dependent upon the path
of the underlying $\{S_t\}_{t\in I}$, $I\subset [0,T]$, hitting a barrier. Our examples will concentrate upon European style
options
$$
\mathbb{E}_{S_0}[\Phi(\{S_t\}_{t\in I})]
$$
with
$$
\Phi(\{S_t\}_{t\in I}) = \mathbb{I}_{A}(\{S_t\}_{t\in I})e^{-rT}(S_T-K)_{+}
$$
a barrier call option, with strike $K>0$, interest rate $r>0$ and $A$ the barrier set. It is assumed
that the initial value of the underlying lies inside (resp.~outside) for knock-out (resp.~knock-in) options. For example, for a discretely monitored
knock-out barrier option, $I=\{t_1,\dots,t_m:0<t_1<\cdots<t_m=T\}$, $t_0=0$,
$d=1$, barrier set $A=\bigotimes_{i=1}^m [a_{t_i},b_{t_i}]$,
the arbitrage price is:
\begin{equation}
\int e^{-rT}(s_{t_m}-K)_{+} \prod_{i=1}^m \bigg\{\mathbb{I}_{[a_{t_i},b_{t_i}]}(s_{t_i})
p(s_{t_i},v_{t_i}|s_{t_{i-1}}v_{t_{i-1}})\bigg\} p(v_0)d(s_{t_1:t_m},v_{t_0:t_m})
\label{eq:barrieroption}
\end{equation}
where $v_{t_i}$ summarizes all the random variables induced by the volatility
process (e.g.~including integrated volatilities) and
it is assumed that the possibly unknown transition densities can be written w.r.t a dominating measure $d(s_{t_1:t_m},v_{t_0:t_m})$.
There are a large number of publications
associated to the Monte Carlo calculation of barrier options, including:
\cite{baldi,glasserman2}; Zhao et al.~(2006)
and for the calculation of the Greeks \cite{bernis,fournie}. 

\subsubsection{A Numerical Example}\label{sec:barriersimulation}

On inspection of (\ref{eq:barrieroption}), it is clear that the estimation
of the barrier option is not a simple task in Monte Carlo integration. For
example, even if we are able to assume that 
\begin{itemize}
\item{the transition densities are known} 
\item{it is possible to simulate from the process} 
\end{itemize}
or if an Euler discretization is adopted, it is still the case that
many paths may yield a zero Monte Carlo estimate: they may knock-out before
the terminal time. A simple remedy to this problem can be found
in \cite{glasserman2}. When the volatility is deterministic, 
the authors
sample from the one-step process, conditioned
so that it does not knock-out. This corresponds to an IS
procedure with weights:
$$
w = \prod_{i=1}^m \bigg\{\int_{[a_i,b_i]} p(s_{t_i}|s_{t_{i-1}},v_{t_{i-1}})ds_{t_i}\bigg\}.
$$
The weights need not be known, but an unbiased estimate of them is required;
see Section \ref{sec:randomweight}, but the point is established in \cite{glasserman2}.
Note, as stated in \cite{glasserman2}, this idea only reduces the variance
of ordinary Monte Carlo in terms of knocking out; it may be sensible to design
an IS strategy that takes into account the nature of the
terminal cost.

The above idea, whilst very effective, is not always useful for large $m$ - the variance of the weights will increase with time. This is illustrated in Table \ref{tab:bsbarrieress},
where a Black-Scholes model is adopted, with $d=1$, $r=0.01$, $\sigma=0.75$, $K=S_0=10.0$, $t_{i}-t_{i-1}=0.5$
and a time-homogenous barrier $A=[5.0,\infty)^{m}$; a down-and-out option. Monte Carlo methods are not needed here, but the idea is to show
how SMC methods help; a more convincing example can be found in Section
\ref{sec:pricingasians}.
The variance of the weights are approximated using the effective sample size
(ESS, see \cite{liu}): 
$$
ESS = \frac{\big(\sum_{i=1}^N w^{(i)}\big)^2}{\sum_{i=1}^N \big(w^{(i)}\big)^2}.
$$
If the simulated samples are dependent, it measures the number of
samples that are independent. Here, it gives us a measure of the variability
of the weights: the closer to $N$ ($=30000$ here), the more the number of `useful' samples.
In Table \ref{tab:bsbarrieress}, it is seen that the variance increases with
$m$; this is well-known (e.g.~the Theorem in \cite{kong}). In some
cases, the algorithm can stabilize; the ideas in the subsequent Sections are still relevant, but are potentially less useful there.

It could be argued that any number of Monte
Carlo enhancement methods could be used to deal with weight degeneracy; however,
none of these approaches are explicitly designed to deal with this. However, SMC methods can deal
with this problem; see Section \ref{sec:pathdegeneracy}, for some associated
drawbacks. 
As noted in Chapter 7 of \cite{glasserman},
the well-known likelihood ratio method \cite{broadie2,gelman} for greek calculation, suffers from exactly this problem. 
As a result, SMC methods will be useful for a wide
variety of option pricing and hedging problems.

\begin{table}\centering
\begin{tabular}{|c|c|c|c|c|c|}
\hline
$m$ & 5  & 10 & 15 & 20 & 25\\
\hline
ESS & 21826.90 & 13389.60 & 8710.91 & 5909.51 & 4139.27\\
\hline
\end{tabular}
\vspace{0.1cm}
\caption{Effective Sample Size for an IS Estimate of a down-and-out Barrier
Option. We used a Black-Scholes model with $r=0.01$, $\sigma=0.75$, $a=5.0$, $b=\infty$.
30000 samples were simulated.}
\label{tab:bsbarrieress}
\end{table}

\section{Sequential Monte Carlo Methods}\label{sec:smcmethods}

\subsection{Sequential Importance Sampling}\label{sec:sir}

We now introduce a general methodology of sequential importance sampling
(SIS), in a particular context. Sequential Monte Carlo techniques can be traced back to at least the 1950's in \cite{hammersley} and \cite{rosenbluth}; see \cite{liu} for a historical review. These ideas have been developed within statistics (e.g.~\cite{doucet2}), physics (e.g.~\cite{jarzynski}), engineering (e.g.~\cite{gordon}) and operations research (e.g.~\cite{glynn}). There is no claim that the reference
list is exhaustive, due to the volume of publications in this field. Much
of this review focuses upon the statistical literature (with numerous exceptions),
since it appears such ideas are not often used in the financial engineering
literature.

The basic idea is as in ordinary IS, the
only real difference being that calculations are performed sequentially.
This sequential formulation allows us both to theoretically understand the
algorithms as well as to derive more advanced versions. The simulation begins
by simulating a collection of samples in parallel and the importance weights
are computed in a sequential manner. These samples, in Section \ref{sec:siralgorithm},
will interact. In the statistics and engineering literature, it is typical to call the samples particles and this terminology is used interchangeably
throughout the paper.

\subsubsection{Formulation}\label{sec:formulation}

Let $\{(E_n,\mathcal{E}_n)\}_{0\leq n \leq m}$ be a sequence of  measurable spaces. It is assumed that it is of interest to
simulate from and compute expectations w.r.t a sequence of related probability measures $\{\pi_n\}_{0\leq
n\leq p}$ on measurable spaces $(E_{[0,n]},\mathcal{E}_{[0,n]})$, 
that is, a sequence of spaces of increasing dimension. Throughout the paper,
$\{\pi_n\}_{0\leq
n\leq p}$ will often be referred to as `targets'. Note, in some scenarios
these targets can be artificial/arbitrary and provide the potential user
with an extra degree of freedom to design the algorithm; this is illustrated in Section \ref{sec:barrier_revisted}.

Introduce a sequence of probability measures of the following
standard form, for $n\geq 0$:
\begin{equation*}
Q_n(x_{0:n}) = M_n(x_n|x_{0:n-1})Q_{n-1}(x_{0:n-1})dx_{0:n}
\end{equation*}
with $M_n:E_{[0,n]}\rightarrow\mathbb{R}^+$ a probability
kernel which we are able to simulate from, $Q_{-1}(\cdot):=1$ and $M_0=\eta_0:E_0\rightarrow\mathbb{R}^+$ a probability density on $E_0$.

The simulation idea, in Figure \ref{fig:gensisalgo}, is then essentially associated to the simple importance
sampling identity:
\begin{eqnarray}
\mathbb{E}_{\pi_n}[h_n(X_{0:n})] & = & \frac{\int_{E_{[0,n]}} h_n(x_{0:n})\big\{\prod_{i=0}^n W_{i}(x_{0:i})\big\}Q_n(x_{0:n})dx_{0:n}}
{\int_{E_{[0,n]}}\big\{ \prod_{i=0}^n W_{i}(x_{0:i})\big\}Q_n(x_{0:n})dx_{0:n}}\label{eq:fksis}\\
W_{i}(x_{0:i}) & = & \frac{\pi_i(x_{0:i})}{\pi_{i-1}(x_{0:i-1})M_i(x_i|x_{0:i-1})}\label{eq:fksisincweight}
\end{eqnarray}
($\pi_{-1}(\cdot):=1$) in the above equation (\ref{eq:fksis}) there is division
by
\begin{equation*}
\int_{E_{[0,n]}} \big\{\prod_{i=0}^n W_{i}(x_{0:i})\big\}Q_n(x_{0:n})dx_{0:n}
\end{equation*}
to ensure that the incremental weights (\ref{eq:fksisincweight}) need
only be known point-wise up to a normalizing constant. The following biased (for finite
$N$, but provably convergent as $N\rightarrow\infty$) estimate
is employed
\begin{eqnarray}
\pi_n^N(h_n)& = & \sum_{l=1}^Nh_n(x_{0:n}^{(l)})w_n^{(l)}\nonumber\\
w_n^{(l)} & = & \frac{\prod_{i=0}^n W_{i}(x_{0:i}^{(l)})}{\sum_{j=1}^N\{\prod_{i=0}^n W_{i}(x_{0:i}^{(j)})\}}\label{eq:importance_weight}
\end{eqnarray}
where $x_{0:n}^{(l)}$ is the $l^{th}$ sample at time $n$.
From herein it is assumed, unless otherwise written, that the incremental weights are the un-normalized versions. The normalizing constant is (abusively)
defined as
$$
Z_n := \int_{E_{[0,n]}}\big\{\prod_{i=0}^n W_{i}(x_{0:i})\big\}Q_n(x_{0:n})dx_{0:n}.
$$

In order to select the $M_n$, a conditionally optimal density is \cite{doucet1}:
\begin{equation}
M_n(x_n|x_{0:n-1}) = \pi_n(x_n|x_{0:n-1})\label{eq:optimalproposal}.
\end{equation}
The proposal is optimal, in terms of minimizing the variance of the incremental
weights, conditional upon $x_{0:n-1}$ (see \cite{doucet} for some
limitations). Algorithms which can be expected to
work well, will attempt to approximate this density; see \cite{doucet1} and the references there-in for details.

There are a large number of extensions of the SIS method.
We list some references here: \cite{chen,delmoral3,doucet,pitt}. In the paper \cite{delmoral3}, the algorithms are combined with
Markov chain Monte Carlo (MCMC) methods; this is discussed later in Section \ref{sec:smcsamplers}.

\begin{figure}
\begin{itemize}
\item{0. Set $n=0$; for each $i\in\mathbb{T}_N$ sample $X_0^{(i)}\sim\eta_0$ and compute $W_0(x^{(i)}_0)$. }
\item{1. Set $n=n+1$, if $n=m+1$ stop, else; for each $i\in\mathbb{T}_N$ sample $X_n^{(i)}|x_{0:n-1}^{(i)}\sim M_n(\cdot |x_{0:n-1}^{(i)})$, compute $W_n(x_{0:n}^{(i)})$ and return to the
start of 1.}
\end{itemize}
\caption{A Generic SIS Algorithm.}
\label{fig:gensisalgo}
\end{figure}

\subsubsection{The Barrier Option Revisited}\label{sec:barrier_revisted}

In the case of the barrier option problem, in Section \ref{sec:barrieroptions},
if it is not possible to simulate from the process, but the transition
densities are known, then we can introduce a process via $Q$. The formula (\ref{eq:barrieroption}) can be approximated using SIS, with the incremental
weight
$$
W_{i}(s_{t_{i-1}:t_i},v_{t_{i-1}:t_i}) = \frac{p(s_{t_i},v_{t_i}|s_{t_{i-1}},v_{t_{i-1}})}{q(s_{t_i},v_{t_i}|s_{t_{i-1}},v_{t_{i-1}})}
$$
$M_i=q$.
Any discretely sampled option pricing problem can be written in this form.
Indeed, in more generality, we may try to incorporate a terminal reward
into the target densities, $\{\pi_n\}$. For example, the optimal importance
density is
$$
Q_m^{s_0}(v_0, s_{t_1:t_m},v_{t_{1}:t_m}) \propto (s_{t_m}-K)_{+}\bigg\{\prod_{i=1}^m \mathbb{I}_{[a_{t_i},b_{t_i}]}(s_{t_i})
p(s_{t_i},v_{t_i}|s_{t_{i-1}}v_{t_{i-1}})\bigg\} p(v_0).
$$
Then, it is sensible to look for optimal sequences of proposals that approximate
this (as in \cite{glasserman2}). 
In theory, the sequence of densities
\begin{eqnarray*}
\pi_n^{s_0,1}(v_0, s_{t_1:t_n},v_{t_{1}:t_n}) & \propto  & \int (s_{t_m}-K)_{+} \bigg\{\prod_{i=1}^m \mathbb{I}_{[a_{t_i},b_{t_i}]}(s_{t_i})
p(s_{t_i},v_{t_i}|s_{t_{i-1}}v_{t_{i-1}})\bigg\}
\times \\ & & 
 p(v_0) d(s_{t_{n+1}:t_m},v_{t_{n+1}:t_m})
\end{eqnarray*}
is a `sensible' path to the optimal IS density; however, they cannot be computed. 
One strategy to circumvent this, in \cite{jasra}, is to introduce a monotonic transformation of the potential $(S-K)_+$, say $g$, at some time, close to the terminal
time in the target density $\pi_n$: 
$$
\pi_n^{s_0,2}(v_0, s_{t_1:t_n},v_{t_{1}:t_n}) \propto g((S_{t_n}-K)_{+}) \bigg\{\prod_{i=1}^n \mathbb{I}_{[a_{t_i},b_{t_i}]}(s_{t_i})
p(s_{t_i},v_{t_i}|s_{t_{i-1}}v_{t_{i-1}})\bigg\} p(v_0).
$$
Note, in many cases $\pi_n^{s_0,1}(\cdot)\leq C_n \pi_n^{s_0,2}(\cdot)$
for $C_n\in(0,+\infty)$, $C_n\geq C_{n+1}$, therefore it is sensible to introduce the potential
function at some time different than 1.
In this case, the estimate
\begin{equation}
\widehat{Z}_m\sum_{i=1}^N\frac{(s_{t_m}^{(i)}-K)_{+}}{g((s_{t_m}^{(i)}-K)_{+})}w_m^{(i)}
\label{eq:estsmcnc}
\end{equation}
can be used. It is described how to approximate the normalizing constant $Z_m$ below.

\subsection{SIR}\label{sec:siralgorithm}

As we saw in Section \ref{sec:barriersimulation}, the SIS method will not
always
work as the time parameter increases. Before continuing, there are related
methods in rare event simulation, termed multi-level splitting (e.g.~\cite{lecuyer1}). These techniques are related to SMC as they are approximations
of multi-level Feynman-Kac formulae \cite{delmoral,cerou};
SMC algorithms are approximations of `standard' Feynman-Kac formulae (as
in \cite{delmoral}). Indeed, SMC algorithms related to splitting are given in \cite{chen}. Since most option pricing problems are not of rare-event
form, we do not discuss the ideas of splitting any further; see \cite{lecuyer1} and the references therein for an introduction.

The formulation is as in Section \ref{sec:formulation}; to simulate from
a sequence of related probability measures on state-spaces of increasing
dimension.
The following resampling scheme is inserted into the SIS algorithm:
At time $n$ of the algorithm 
$N$ particles are sampled, with replacement, from the current set of particles according to some stochastic rule with the property
\begin{equation}
\mathbb{E}[N^i_n|x_{0:n}^{(i)}] = Nw_{n}^{(i)}
\label{eq:resamplecrit}
\end{equation}  
where $N^i_n$ is the number of replicates of the $i^{th}$ particle at time
$n$. The most basic way this can be achieved is by resampling the particles
according to the normalized weights $\{w_n^{(i)}\}_{1\leq i\leq N}$.
There are a variety of superior methods for resampling the particles: 
residual, stratified, systematic etc; a full description can be found in
\cite{doucet2}. The systematic method is used here. This sequential importance sampling with resampling, is termed sequential importance sampling/resampling (SIR) (see \cite{delmoral,delmoral1,doucet2}). The simulated paths are no longer independent,
but there is a well-established convergence theory; see \cite{delmoral}.
In addition, there is a theoretical advantage to resampling the particles;
the asymptotic variance in the central limit theorem \cite{delmoral} can be upper-bounded, uniformly in time, with resampling; this is not necessarily the case otherwise - see \cite{chopin}. 

The algorithm is given in Figure \ref{fig:gensiralgo}. The particles
can be resampled at any time step of the algorithm, but it is best to do
this when the weights are very variable; e.g.~when the ESS drops below a
pre-specified threshold. 
This is theoretically valid, as established in
\cite{delmoral2}. The reason for the above is that if the particles are resampled too often, then there are too many sampled paths which have been replicated - the paths degenerate; see Section \ref{sec:pathdegeneracy} for further details. In addition, if the optimal proposal (\ref{eq:optimalproposal})
can be used, then it is best to sample after the resampling operation has
occurred. This will increase the number of unique samples and lower the
variance in estimation.

To estimate the normalizing constants, use
\[
\widehat{Z_{n}}=\prod\limits_{j=1}^{r_{n-1}}\widehat
{\frac{Z_{k_{j}}}{Z_{k_{j-1}}}},
\]
with
\begin{equation}
\widehat{\frac{Z_{k_{r}}}{Z_{k_{r-1}}}}=\frac{1}{N}\sum_{i=1}^{N}\prod\limits_{j=k_{r-1}+1}^{k_{r}}W_{j}\left(
x_{0:j}^{\left(  i\right)  }\right)
\label{eq:estimatorrationormalizingconstants2}%
\end{equation}
where $k_{0}=0$, $k_{j}$ is the $j^{\text{th}}$ time index at which one
resamples for $j>1$. The number of resampling steps between $0$ and $n-1$ is denoted $r_{n-1}$ and we set $k_{r_{n}}=n$ (to ensure that the final term includes $Z_n$ in the numerator, as is required for correctness). 

\begin{figure}
\begin{itemize}
\item{0. Set $n=0$; for each $i\in\mathbb{T}_N$ sample $X_0^{(i)}\sim\eta_0$ and compute $W_0(x^{(i)}_0)$. }
\item{1. Decide whether or not to resample, and if this is performed, set
all weights to $w_n^{(i)}=1/N$ and proceed to step 2. If no resampling occurs, the weights are as in equation \eqref{eq:importance_weight}.}
\item{2. Set $n=n+1$, if $n=m+1$ stop, else; for each $i\in\mathbb{T}_N$ sample $X_n^{(i)}|x_{0:n-1}^{(i)}\sim M_n(\cdot |x_{0:n-1}^{(i)})$, compute $W_n(x_{0:n}^{(i)})$ and return to the
start of 1.}
\end{itemize}
\caption{A Generic SIR Algorithm.}
\label{fig:gensiralgo}
\end{figure}

\subsubsection{SMC Samplers}\label{sec:smcsamplers}

A useful algorithm related to SIR is termed SMC samplers \cite{delmoral3}.
This method is designed to sample from a sequence of target distributions
on a common space; i.e.~$\{\pi_n\}_{0\leq n\leq p}$ are probabilities on
$(E,\mathcal{E})$. However, due to a difficulty described below, the algorithm
samples from a sequence of distributions of increasing dimension. 
The marginal of each new density is the one of interest.

Suppose we initialize an IS algorithm with $N$ particles $\{X_0^{(i)}\}_{1\leq
i\leq N}$ sampled
according to some initial density, $\eta_0$, and weight w.r.t $\pi_0$.
Further, suppose that the IS works well; 
this can be achieved by making $\pi_0$ very simple, e.g. $\eta_0=\pi_0$.
Now, under the assumption that $\pi_0$ and $\pi_1$ are not significantly
different, it might be expected that we can construct a Markov kernel $K_{1}(x_0,x_1)$
so as to move the particles from regions of high density of $\pi_0$
to the corresponding regions of $\pi_1$. In such a case, the importance weight would be:
\begin{equation}
w_1(x_1^{(i)}) = \frac{\pi_1(x^{(i)}_1)}{\int_E\eta_0(x_0)K_1(x_0,x_1^{(i)})dx_0}.
\label{eq:smcsampproblem}
\end{equation}

Equation (\ref{eq:smcsampproblem}) presents some major difficulties:
\begin{itemize}
\item{In many cases the integral in the denominator (\ref{eq:smcsampproblem})
cannot be computed.}
\item{In some cases, $K_1$ is not known point-wise.}
\end{itemize}
For the above reasons, it appears that such a weighting scheme seems destined
to fail. For example, even when there is the computational power
to approximate the integral 
in the denominator of (\ref{eq:smcsampproblem}):
$$
\frac{1}{N}\sum_{j=1}^N\eta_0(x_0^{(j)})K_1(x_0^{(j)},x_1^{(i)})
$$
the second point will prohibit its application.

The solution is to
modify the problem to a more familiar setting. Introduce a sequence
of auxiliary probability measures $\{\widetilde{\pi}_n\}_{0\leq n \leq p}$
on state-spaces of increasing dimension $(E_{[0,n]},\mathcal{E}_{[0,n]})$,
such that they admit the $\{\pi_n\}_{0\leq n\leq p}$ as marginals. 
The following sequence of auxiliary densities is used (see \cite{jarzynski}
and \cite{neal}):
\begin{equation}
\widetilde{\pi}_n(x_{0:n}) = \pi_n(x_n)\prod_{j=0}^{n-1}L_{j}(x_{j+1},x_j)
\label{eq:smcsampaux}
\end{equation}
where $\{L_{n}\}_{0\leq n \leq p-1}$ are a sequence of Markov kernels that
act backward in time and are termed backward Markov kernels. The algorithm
samples forward using kernels $\{K_n\}$.
The choice of
backward kernels is made as the incremental weights are
\begin{equation}
W_n(x_{n-1:n}) = \frac{\pi_n(x_n) L_{n-1}(x_n,x_{n-1})}{\pi_{n-1}(x_{n-1})K_n(x_{n-1},x_n)}
\quad n\geq 1\label{eq:smc_inc_weight1}
\end{equation}
which allows for a fast computation and avoids a path degeneracy effect (see
Section \ref{sec:pathdegeneracy}).
It is clear that (\ref{eq:smcsampaux}) admit the $\{\pi_n\}$ as marginals,
and hence, if one sequentially samples from $\{\widetilde{\pi}_n\}$ we are left with a problem that is the same as for SIR. That is, one uses 
the algorithm in Figure \ref{fig:gensiralgo} with 
the kernel $K_n$
as a proposal (instead of $M_n$) and incremental weight \eqref{eq:smc_inc_weight1}. A
discussion of the optimal choice of backward kernels can be found in \cite{delmoral3}.

\subsubsection{The Barrier Option Revisited}\label{sec:barriersimossir}

Two examples of SIR algorithms are now presented.
The first, is a straightforward modification of the SIS method demonstrated
in Section \ref{sec:barriersimulation}. In this case the target densities are
$$
\pi_n^{s_0}(v_0,s_{t_1:t_m}) \propto 
\bigg\{\prod_{i=1}^n \mathbb{I}_{[a_{t_i},b_{t_i}]}(s_{t_i})
p(s_{t_i}|s_{t_{i-1}})\bigg\},
$$
with importance densities as in Section \ref{sec:barriersimulation}.
As a result, this approach is as in \cite{glasserman2},
except with resampling. The second idea is as in \cite{jasra}: include the function $(S-K)$ in the target. This is to reduce the variance
in estimation, when considering the call option. The function used is
$$
g(S-K) = |(S-K)|^{\kappa},
$$
$\kappa$ is referred to as a temperature parameter. See \cite{jasra}
for a justification of this choice of $g$.

\begin{figure}[h]\centering
\subfigure[No Potential.]
{\includegraphics[width=0.48\textwidth,height=6cm]{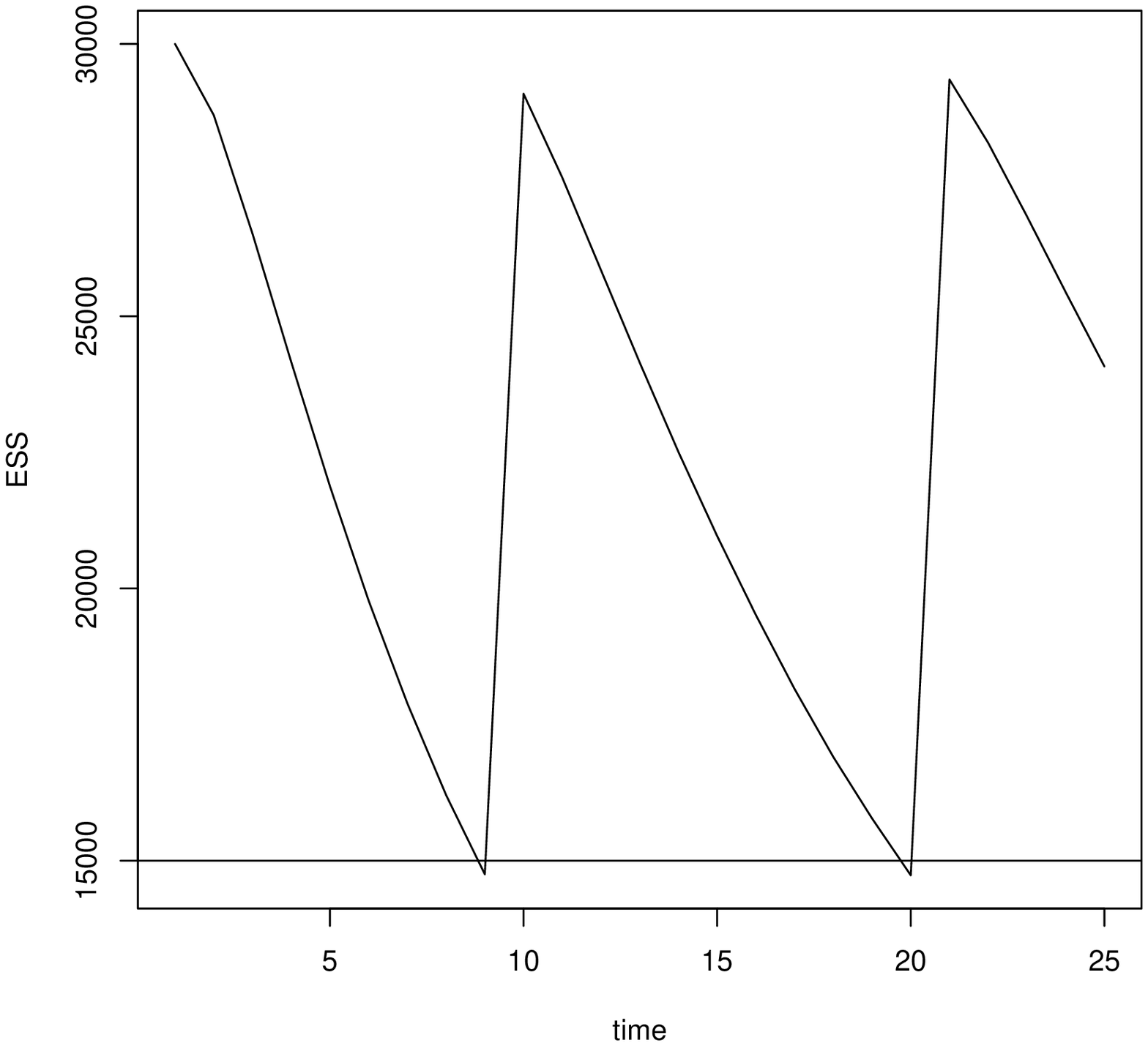}}
\subfigure[Potential.]
{\includegraphics[width=0.48\textwidth,height=6cm]{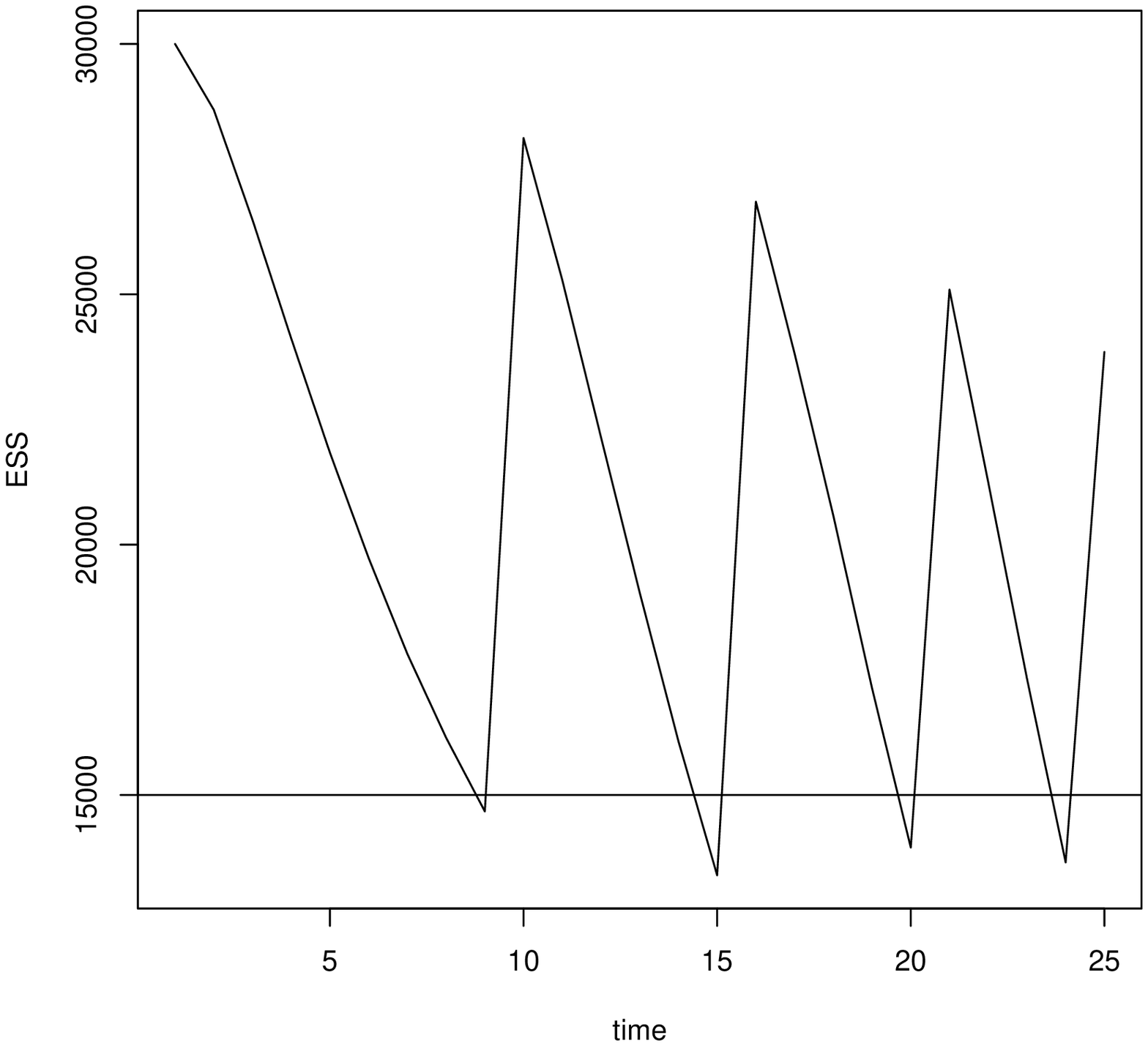}}
\caption{Controlling the Variance of the Weights. We used an SIR algorithm
to apply the method of \cite{glasserman2} to a simple Black-Scholes model (left). In the right plot we used the approach of \cite{jasra}.
The settings are as in Section \ref{sec:barriersimulation} and there are $m=25$ time steps of
the algorithm and 30000 particles are simulated.}
\label{fig:opmc1}
\end{figure}

In Figure \ref{fig:opmc1} (a) the performance of the first SIR method can
be seen. In the Figure, we have used the Black-Scholes model with $m=25$
and $N=30000$ particles; the particles are resampled when the ESS drops below
15000 particles. The algorithm exhibits very stable variance properties,
resampling only 2 times. As a result, and in comparison to the results of
Table \ref{tab:bsbarrieress}, the algorithm is much more useful for the estimation
of the down-and-out barrier option. Note that there is a marginal increase
of the overall CPU time. The CPU time is less than 10 seconds on a Pentium 4 2.4 GhZ machine, coded in C++; all code is available upon request from the author.

In Figure \ref{fig:opmc1} (b) the performance of the second SIR algorithm
can be seen. In this case the potential is introduced at time 10, with $\kappa=0.08$.
At subsequent time steps, this value increases by 0.045. These values are
set by some prior tuning, but it is possible to do this adaptively; see \cite{jasra1}.
The issues here are:
\begin{itemize}
\item{When to introduce the potential}
\item{How fast the temperature should be increased.}
\end{itemize}
In general, the potential could be introduced quite early in the simulation.
This would allow the samples to adapt to the potential, but at the cost of
increasing the variability of the weights. In practice, we have used between
one-third and two-thirds of the overall algorithm time-length. The second issue is
also important: to reduce the variability of (\ref{eq:estsmcnc}), the temperature
should be (at the final time-step) bigger than 1. However, if the temperature
increases too rapidly, then algorithm has to resample too often.
The stability of
the weights is not as good as in Figure \ref{fig:opmc1} (a). This is due
to the introduction of the potential function. However, let us consider the
actual quality in estimation of the down-and-out option, a significant improvement
can be seen. Both approaches are run 25 times, with 30000 particles, resampling
threshold of 15000, as well as computing the analytic approximation in \cite{broadie1};
the estimates of the options ($\pm 2$ standard deviations across the repeats) are $76.81 \pm 13.81$ and $6.03 \pm 0.43$, with analytic approximated value of 6.16. This illustrates
a well-known fact about IS: the samples have to be pushed towards regions
of importance, in terms of the functional of interest. The monitoring of
the variance of the weights may not be enough to yield sensible estimates
for option pricing.

\subsection{SMC Methods for Sensitivities}

We now review some SMC methods for estimating derivatives of expectations.
This can be useful for calculating the Greeks, especially when the transition
densities are not known and Euler approximations are adopted via the likelihood
ratio method. There
is a growing literature on Malliavin calculus approaches \cite{bernis,fournie}, even in the case of quite sophisticated SV models \cite{benth};
these problems could be dealt with using the SMC approaches in the other
Sections. Indeed the rates of convergence, in comparison to the likelihood
ratio method, are faster \cite{detemple}, although in many cases the
Malliavin weight is not known.

\subsubsection{Path Degeneracy}\label{sec:pathdegeneracy}

In option pricing, most Greeks which are not calculated through Malliavin
methods are of the form
\begin{equation}
\Lambda(\Phi,s_0) = 
\nabla\bigg[\int \Phi(s_{t_1:t_m}) \bigg\{\prod_{i=1}^m p(s_{t_i},v_{t_i}|s_{t_{i-1}},v_{t_{i-1}})\bigg\}
p(v_0)d(s_{t_1:t_m},v_{t_0:t_m})\bigg]\label{eq:optsensitivity}.
\end{equation}
Monte Carlo methods for computing this quantity can be found in Chapter 7
of \cite{glasserman}, see also \cite{asmussen}. It is explicitly assumed that the transition density
is known and upper bounded, and that the derivatives of this quantity can be calculated analytically.

One of the problems of estimating (\ref{eq:optsensitivity}) using IS methods, is that as seen in
Section \ref{sec:barriersimossir}, resampling
is often required to control the variance of the weights. However, a key
problem for estimating (\ref{eq:optsensitivity}) is that the integral is on a path space.
For example, approaches which calculate
expectations (option prices) can 
use the same set of particles to estimate the sensitivities (Greeks)
\cite{cerou1}. The difficulty is the following: the resampling mechanism, whilst
necessary to control the variance of the weights, induces a path degeneracy
effect. There may be many unique particles at the current time, but going
backward in time yields paths which coalesce to the same parents.
That is to say, that $x_{0:n-1}$ are not re-simulated at time $n$. Hence,
due to resampling, many of the $x_{0:n-L+1}$ ($L>0$) will be the same across the simulated samples.
In other words SMC algorithms are only useful for calculating expectations
of the form:
$$
\int_{E_{[n-L,n]}} h(x_{n-L:n})\pi_n(x_{n-L:n})dx_{n-L:n}
$$
for some \emph{fixed} lag $L>0$. See \cite{andreiu1} and \cite{delmoral} for more details.
It should be noted that it will still be possible to compute the sensitivities
of a wide-class of option prices.

\subsubsection{Marginal Approximations}

In this Section it is shown how to estimate the Greeks, when using the same
set of particles to estimate the option price. The method can only be used if there
is a Markovian dependence of the $(S_{t_n},V_{t_n})\in E_n$. For simplicity
$v_0$ is dropped, but can straight-forwardly be re-introduced.

The idea is associated to the likelihood
ratio method; assume the pay-off function is of the form
$$
\Phi(S_{t_1},\dots,S_{t_m}) = \prod_{i=1}^m \widetilde{\Phi}(S_{t_i})
$$
note that there are wide class of options that can be written this way.
Assume that only the process depends upon a parameter $\theta\in\Theta$. Then the sensitivity is, assuming the validity of interchanging integration and differentiation (e.g.~\cite{lecuyer})
$$
\Lambda_{\theta}(\Phi,s_0) = \int \bigg\{\prod_{i=1}^m \widetilde{\Phi}(S_{t_i})\bigg\}
\frac{\partial}{\partial
\theta}\bigg\{\prod_{i=1}^m  p_{\theta}(s_{t_i},v_{t_{i}}|s_{t_{i-1}},v_{t_{i-1}})\bigg\}  d(s_{t_1:t_m},v_{t_1:t_m})
$$
where $\theta$ is the parameter of interest.
To obtain
the sensitivity of interest we present the following result, whose proof is housed in the appendix.

\begin{proposition}\label{prop:recursion}
Let, for $n\geq 2$
\begin{eqnarray}
\widetilde{\Lambda}_{n}(dx_{n}) & := &
\bigg[\int_{E_{n-1}}\widetilde{\Lambda}_{n-1}(dx_{n-1})p_{\theta}(x_n|x_{n-1})\widetilde{\Phi}(s_{t_n})
\bigg]dx_n + \label{eq:sensitivityrecursion}\\ & & 
\bigg[\int_{E_{n-1}} 
\Pi_{n-1}(dx_{n-1})
\frac{\partial}{\partial\theta}\{p_{\theta}(x_n|x_{n-1})\}
\widetilde{\Phi}(s_{t_n})\bigg]dx_{n}\nonumber
\end{eqnarray}
where
$$
\Pi_{n}(dx_{n}) = \bigg[\int_{E_{n-1}}\Pi_{n-1}(dx_{n-1})p_{\theta}(x_n|x_{n-1})\widetilde{\Phi}(s_{t_n})
\bigg]dx_n
$$
$x_{n}=(s_{t_n},v_{t_n})$ and
\begin{eqnarray*}
\widetilde{\Lambda}_1(dx_1) & = & \bigg[\frac{\partial}{\partial\theta}\{
p_{\theta}(x_1|x_{0})\}\widetilde{\Phi}(s_{t_1})
\bigg]dx_1\\
\Pi_{1}(dx_{1}) & = & \bigg[ p_{\theta}(x_1|x_0)\widetilde{\Phi}(s_{t_1})
\bigg]dx_1.
\end{eqnarray*}
Then, $\widetilde{\Lambda}_m(1)=\Lambda_{\theta}(\Phi,s_0)$.
\end{proposition}

In the case of the Greeks $\Delta$ and $\Gamma$,
the second expression on the RHS (\ref{eq:sensitivityrecursion}) is not needed,
due to the dependence of only the first term in the product 
$\prod_{i=1}^{n-1}p_{\theta}(x_i|x_{i-1})$ on $s_0$.

The approach below is a special case of that detailed in \cite{poyiad}
(see also \cite{poyiad1}). The objective is to recursively approximate the measures $\{\widetilde{\Lambda}_n\}$
and $\{\Pi_n\}$. The initialization can
be achieved by IS. Then, given an
empirical approximation at time $n-1$ of $\Lambda_{n-1}$, say:
\[
\sum_{i=1}^{N}W_{n-1}^{\left(  i\right)  }\delta_{x_{n-1}^{(i)}}(dx_{n-1})
\]
and of $\Pi_{n-1}$,
with different weights $\widetilde{W}_{n-1}$, then the following is a point-wise approximation of the signed density
\[
 \widetilde{\Lambda}_{n}^{N}(x_{n})=
 \widetilde{\Phi}(s_{t_n})\bigg[
 \sum_{i=1}^{N}W_{n-1}^{\left(  i\right)
}p_{\theta}(x_n|x_{n-1}^{(i)})
+\sum_{i=1}^N \widetilde{W}_{n-1}^{(i)}\frac{\partial}{\partial\theta}\{p_{\theta}(x_n|x_{n-1}^{(i)})\}
\bigg].
\]
Assuming that new particles $\left\{  X_{n}^{(i)}\right\}
$ are sampled from some density $\Psi_{n}^{N}(x_{n})$, then using IS, the following approximation is obtained
\[
\widetilde{\Lambda}_{n}^{N}(dx_{n})=\sum_{i=1}^{N}W_{n}^{\left(  i\right)  }\delta
_{x_{n}^{(i)}}(dx_{n})
\]
where
\begin{equation}
W_{n}^{\left(  i\right)  }=\frac{\widetilde{\Lambda}_{n}^{N}(x_{n}^{(i)})}%
{N\Psi_{n}^{N}(x_{n}^{\left(  i\right)  })}.\label{eq:mpfrecursion}%
\end{equation}
Note
$$
\widetilde{W}_n^{(i)} = \frac{\sum_{j=1}^N \widetilde{W}_{n-1}^{(j)}p_{\theta}(x_n^{(i)}|x_{n-1}^{(j)})}{N\Psi_{n}^{N}(x_{n}^{\left(  i\right)  })}.
$$

The computational complexity of computing the weights is $O(N^{2})$. This is required,
to avoid the path degeneracy problem. It should be noted that an asymptotically biased approach,
which is of $O(N)$ could, in some cases, be used from \cite{olsson}, with the identity
$$
\frac{\partial}{\partial \theta} \int h(x)\pi_{\theta}(x)dx
=
\int h(x)\frac{\partial}{\partial \theta}\log\{\pi_{\theta}(x)\}\pi_{\theta}(x)dx.
$$
See \cite{olsson} for further details. We remark, also, that the work
of \cite{delmoral4,delmoral5} can also be used, in some contexts, for greek calculation, but do not review these new techniques here.

\subsubsection{Barrier Option Revisited}\label{sec:barrier_sen_smc}

We return to the Example in Section \ref{sec:barriersimulation}. We calculate
the $\Delta$ and the $\mathcal{V}$ using the recursion in (\ref{eq:sensitivityrecursion}).
The model settings were as in Section \ref{sec:barriersimulation}. The algorithm
is run with 10000 particles, 25 times which took 30 hours; i.e.~1 hour 20
min to run the algorithm once. The $\Delta$ was estimated as $0.22\pm
1.14$ and the $\mathcal{V}$ was $-29.7\pm 138.34$. Note that the error is $\pm$ 2 standard deviations for the 25 repeats. 
The ratio of the estimate to the $\pm$ 2 standard deviations are approximately 1/5 and perhaps too 
high; reducing the variability is the subject of current research.
The CPU time is not substantial and shows that the given recursions
are potentially useful for at computing $\Delta$ and $\Gamma$, but other
path-based sensitivities may require biased estimates as in \cite{olsson}.

\subsection{SMC Methods for Continuous-Time Processes}

In the context of continuous-time processes, there are discrete-time SMC algorithms which can approximate the continuous-time Feynman-Kac formulae
with no approximation-bias (see also \cite{rousset}).  

\subsubsection{The Random Weight}\label{sec:randomweight}

An important comment, as realized by \cite{rousset1}, on IS is the following. It is not necessary to know
the weight exactly; only 
to find an unbiased estimate of it.
More exactly, let $n=0$ for simplicity. Suppose it is possible to find some function $\vartheta_n:E_n\times F_n\rightarrow\mathbb{R}^+$
and a probability $\rho_n$ so that:
$$
w_n(x) = \mathbb{E}_{\rho_n(x,\cdot)}[\vartheta_n(x,U)].
$$
then
\begin{eqnarray*}
\pi_n(h_n) & = & \frac{1}{Z_n}\int_{E_n}h_n(x)w_n(x)\eta_n(x)dx\\
& = & \frac{1}{Z_n}\int_{E_n}f_n(x)\bigg\{\int_{F_n}\vartheta_n(x,u)\rho_{n}(x,u)du\bigg\}\eta_n(x)dx.
\end{eqnarray*}
As a result, we may perform joint IS by sampling from $\rho_n\times\eta_n$.
This comment is extremely useful for option pricing; it has already adopted,
by at least \cite{glasserman2}. Moreover, it can be useful for calculating expectations w.r.t probability laws of L\'evy processes.

\subsubsection{Barrier Option Revisited}

The ideas here are based upon the random weight, and the exact simulation
of diffusions methodology \cite{beskos,fearnhead}. In many cases,
the option price can be written as an expectation w.r.t the transition
densities, but that the densities are not available up-to a constant or cannot be simulated. However, it may be that
transition laws are known indirectly e.g.~in terms of their characteristic function; see \cite{glasserman3} for some solutions.

Consider (\ref{eq:barrieroption}) with deterministic volatility
and the underlying is a diffusion satisfying the SDE
$$
dS_t = \xi(S_t) dt + dW_t
$$
with $\xi=\nabla\Xi$ and some additional assumptions in \cite{fearnhead}. Then the transition is
\begin{eqnarray}
p(s_{t_i}|s_{t_{i-1}}) & = & \phi(s_{t_i};s_{t_{i-1}},(t_i-t_{i-1}))\exp\big\{\Xi(S_{t_i}) - \Xi(S_{t_{i-1}} - l(t_i-t_{i-1})) \big\}
\times \nonumber\\ & & 
\mathbb{E}_{\mathbb{W}_{s_{t_{i-1}}}^{s_{t_{i-1}}}}
\bigg[\exp\big\{-\int_{[0,T]}\zeta(W_s)ds\big\}\bigg]
\label{eq:transitiondensity}
\end{eqnarray}
where the expectation is taken w.r.t the law of a Brownian bridge starting
at $s_{t_{i-1}}$, ending at $s_{t_i}$ and the other parameters/functions
are known exactly; see \cite{fearnhead} for details. Therefore, we can use an SIR algorithm with an incremental weight
$$
\frac{p(s_{t_i}|s_{t_{i-1}})}{q(s_{t_i}|s_{t_{i-1}})}
$$
that is not known exactly. However, it is detailed in both \cite{beskos}
and \cite{fearnhead} how the expectation in \eqref{eq:transitiondensity}
can be estimated unbiasedly; the
random weight idea can be applied. In our experience the estimation of 
$$
\mathbb{E}_{\mathbb{W}_{s_{t_{i-1}}}^{s_{t_{i-1}}}}
\bigg[\exp\big\{-\int_{[0,T]}\zeta(W_s)ds\big\}\bigg]
$$
through the methodology in \cite{fearnhead}, can lead to substantial increases
in variance of the weights. In high-dimensional cases it may be difficult to use this idea, without some extra
variance reduction methods. A similar idea can be found in \cite{rousset}
when approximating
$$
\mathbb{E}^{\mathbb{W}}_x \bigg[ g(W_T) \exp\bigg\{ \int_{[0,T]}U(W_s)ds \bigg\} \bigg].
$$
See \cite{papas} for a review of related issues.

\section{An SMC Method for Pricing Asian Options}\label{sec:pricingasians}

In the following Section we present a method to approximate the value of
a fixed strike, arithmetic Asian option, when the underlying follows the system
\begin{eqnarray*}
dY_t & =  & \mu dt + V_t dW_t\\
dV_t & = & -V_t dt + dU_t
\end{eqnarray*}
where $Y_t$ is the log price, $\{W_t\}_{t\in[0,T]}$ is a Brownian motion and $\{U_t\}_{t\in[0,T]}$
is a pure-jumps L\'evy process, such that the marginal of $V_t$ is a Gamma
distribution; see \cite{bns} and the appendix for details. This latter model has been
shown to fit, to an extent, the dynamics of real price data. 

This problem is particularly difficult for Monte Carlo methods, even in the
Black-Scholes case (see, however, \cite{jourdain} for some work on
the exact approximation with continuously monitored average). 
The reason is due to the path-based nature of the payoff, which can contribute
a substantial variance in the estimate. Indeed, given the methodology explained
thus far, it is not straightforward to use SMC methods for exactly the previous
reason; see Section \ref{sec:pathdegeneracy}. Some sensible solutions (to
the form of the pay-off), in the Black-Scholes scenario, can be found in \cite{glasserman1} and \cite{kemna}. Also, note that efficient techniques for partial differential equations (PDEs) can be found in  \cite{vecer,vecer1}.
These methods appear to be adaptable to our case, but would lead to solving
a partial integro-differential equation which is not always simple and accurate; see \cite{cont} for some ideas.

\subsection{Strategy}
The idea is to try, as in the PDE methods above, to reduce the dimension
of the problem. In essence, we seek to sample from the law of the sum of
the underlying at each monitoring period. The problem is that this is only known in terms
of an intractable integral. However, it is shown that it is possible, using
SMC methods to sample from the optimal importance density, such that it is
a marginal on an extended state-space.

In our case, ignoring the discount term, the value of the option is
$$
\int \bigg(\frac{1}{m}\sum_{i=1}^m e^{y_{t_i}} - K\bigg)_{+}
\bigg\{\prod_{i=1}^m p(y_{t_i}|y_{t_{i-1}},v_{1:t_i})p(v_{t_i})\bigg\} d(y_{t_1:t_m},v_{t_1:t_m})
$$
where $v_{t_{0}}$ is known and suppressed from the notation and
$v_{t_i}$ is a 2-dimensional Poisson process on $[0,\lambda\nu(t_i-t_{i-1})]
\times[0,1]$. 
Making the transform,
$s_{t_i}=\exp\{ y_{t_{i}} \}$, and then
\begin{eqnarray*}
S_{t_i} & = & \nu_{t_i} - \nu_{t_{i-1}}\quad i\geq 2\\
S_{t_1} & = & \nu_{t_1}
\end{eqnarray*}
yields the option price
$$
\int \bigg(\frac{\nu_{t_m}}{m} - K\bigg)_{+}
\bigg\{\prod_{i=1}^m \varphi_{\nu_{t_{i-1}}}\{\nu_{t_i};\tilde{\mu}(\nu_{t_{i-2}:t_{i-1}}),
\tilde{\sigma}_i(v_{1:t_i})\}p(v_{t_i})\bigg\} d(\nu_{t_1:t_m},v_{t_1:t_m})
$$
where $\varphi_{\nu_{t_{i-1}}}\{\nu_{t_i};\tilde{\mu}(\nu_{t_{i-2}:t_{i-1}}),\sigma(v_{1:t_i})\}$
is the shifted log-normal density with location parameter
$$
\tilde{\mu}(\nu_{t_{i-2}:t_{i-1}}) = \log(\nu_{t_{i-1}}-\nu_{t_{i-2}}) + \mu(t_{i}-t_{i-1})
$$
scale parameter as in \cite{bns}, $\nu_{t_{-1}}=0$, $\nu_{t_{0}}=S_0$ and see the Appendix
for the definition of $\tilde{\sigma}$.
If there was not a path dependence (see Section \ref{sec:pathdegeneracy}) on the integrated volatilities $v_{t_1:t_m}$,
this transformation would greatly improve
SMC methods. This is because one needs only estimate an integral w.r.t the
marginal $\nu_{t_m}$.
However, the problem is recast
into a similar case as in Section \ref{sec:barriersimossir}; as a result
the method detailed below, is restricted to the
case where $m$ is not large ($m\leq 24$).

\subsection{Simulations}

\subsubsection{SMC Method}
The SMC approach is similar to the second method in Section \ref{sec:barriersimossir}.
The difference is as follows. In prior simulations it was found the temperature
$\kappa$ could not be too large ($\kappa< 0.5$) to avoid a substantial weight degeneracy effect. This is
when introducing the potential in the middle of the simulation.
If the potential function
was introduced very early, the path degeneracy effect occurred; despite the
fact that the temperature is high, which can reduce the variance in estimation, the algorithm resampled too often to yield
poor estimates.

To allow the temperature parameter
to reach 1, an SMC sampler was adopted. That is, once time $m$ is reached
the SMC sampler simulated from
$$
\pi_n(\nu_{t_1:t_m},v_{t_1:t_m}) \propto \bigg|\frac{\nu_{t_m}}{m} - K\bigg|^{\tilde{\kappa_n}}
\bigg\{\prod_{i=1}^m \varphi_{\nu_{t_{i-1}}}\{\nu_{t_i};\tilde{\mu}(\nu_{t_{i-2}:t_{i-1}}),
\tilde{\sigma}_i(v_{1:t_i})\}p(v_{t_i})\bigg\}
$$
for $\kappa<\tilde{\kappa}_1<\cdots<\tilde{\kappa}_p=1$. Here $\kappa$ is the final temperature
of the SIR algorithm.
In all cases, $p=20$ and the $\kappa$'s
increase at a constant amount; see \cite{jasra} for automatic schemes. The
initial SIR is used for computational efficiency; it is less costly to run
this algorithm than SMC samplers. Note that the estimate of the option price
is as in (\ref{eq:estsmcnc}).

For SMC samplers, the particles are simulated according to an MCMC kernel,
$K_n$, of invariant distribution
$\pi_n$; details are given in the appendix.
The backward kernel used in the SMC sampler (see Section \ref{sec:smcsamplers}),
is
$$
L_{n-1}(x_n,x_{n-1}) = \frac{\pi_n(x_{n-1})K_n(x_{n-1},x_n)}{\pi_n(x_{n})}
$$
where $x_n=(\nu_{t_n},v_{t_n})$, 
is such that the incremental weights at time $n$ are
\begin{equation}
W_n(x_{n-1}) = \frac{\pi_n(x_{n-1})}{\pi_{n-1}(x_{n-1})}\label{eq:smcsampler_weight}.
\end{equation}
The choice of backward kernel is quite popular in the literature and, in our experience, often works well in practice. The procedure is summarized
in Figure \ref{fig:asian_smc}.

\begin{figure}
\begin{itemize}
\item{ 0. Use an SIR algorithm (as in Figure \ref{fig:gensiralgo}) to sample
from the sequence of densities:
$$
\pi_n(\nu_{t_1:t_n},v_{t_1:t_n}) \propto \bigg|\frac{\nu_{t_n}}{m} - K\bigg|^{\kappa_n}
\bigg\{\prod_{i=1}^n \varphi_{\nu_{t_{i-1}}}\{\nu_{t_i};\tilde{\mu}(\nu_{t_{i-2}:t_{i-1}}),
\tilde{\sigma}_i(v_{1:t_i})\}p(v_{t_i})\bigg\}
$$
 with $n\in\mathbb{T}_m$, $\{\kappa_n\}_{1\leq n\leq m}$ given ($0 \leq
\kappa_1<\cdots<\kappa_m<1$). The process
densities are used as the proposals.}
\item{1. Given the samples from $\pi_n(\nu_{t_1:t_m},v_{t_1:t_m})$ use SMC samplers (as in Section \ref{sec:smcsamplers}) to sample from 
$$
\pi_n(\nu_{t_1:t_m},v_{t_1:t_m}) \propto \bigg|\frac{\nu_{t_m}}{m} - K\bigg|^{\tilde{\kappa}_n}
\bigg\{\prod_{i=1}^m \varphi_{\nu_{t_{i-1}}}\{\nu_{t_i};\tilde{\mu}(\nu_{t_{i-2}:t_{i-1}}),
\tilde{\sigma}_i(v_{1:t_i})\}p(v_{t_i})\bigg\}
$$
with $n\in\mathbb{T}_p$, $\kappa_{m}=\kappa<\tilde{\kappa}_1<\cdots<\tilde{\kappa}_p=1$. Note
that this is a sequence of densities on the same space, as opposed to the
sequence in step 0.~which samples from one of increasing dimension. The 
SMC sampler uses the MCMC algorithm in the appendix and has incremental weight
\eqref{eq:smcsampler_weight}.}
\end{itemize}
\caption{SMC Method to Approximate Arithmetic Asian Options.}
\label{fig:asian_smc}
\end{figure}

\subsubsection{Importance Sampling Method}

To compare the method above, the approach of \cite{glasserman1} is modified.
The main difference is to simulate the integrated volatility exactly, and
to use the smallest value (over each interval) in the numerical search for
a change of drift. In this scenario, it was found that there was not always a unique root; this is in less than 1\% of the cases in the results below.

\subsubsection{Numerical Results}

The algorithms are run 50 times, for similar CPU time; this was less than
150 seconds. $m=12$, $\lambda=1$, $t_{i}-t_{i-1}=1$ ($\forall
i\in\mathbb{T}_m$), $\mu=0.07$,
$\nu=0.5$; these values, when relevant, are consistent with the parameter estimates that were reported in \cite{griffin}. The SMC method was run with 
10000 samples, and in the SIR part of the algorithm the potential was introduced
at time 6, with $\kappa=0.2$; this increased by  0.035 at each time step, until time 12 was reached. The process densities were used for the proposal densities.
The MCMC moves, in the appendix had acceptance
rates between 0.3-0.5 and this was quite a reasonable performance.
The IS method was run with only 4000 samples.
The higher CPU time, for IS, is due to the fact that a bisection method is run for each sample (which was allowed a maximum of 300 steps).

The variance reduction factors, across the multiple runs, for 5 different settings of $S_0$ and $K$ can be observed in Table \ref{tab:iscomp}. 
Note that whilst the option can be standardized for $S_0$, our aim is simply
to show that the conclusions of this experiment are relatively invariant
to the parameters used.
In the Table, a substantial improvement
for the same CPU times can be observed. It should be noted, however, for
larger $\nu$ (in the latent process) both algorithms do not work well. The latter parameter will increase the variability of the volatility process, and it is not clear how this can be dealt with. Note the results here should
not be contrasted with (slightly disappointing) those in Section \ref{sec:barrier_sen_smc};
this technique differs substantially from the ideas there.

\begin{table}
\centering
\fbox{%
\begin{tabular}
[c]{c|c}%
\textbf{Settings} & \textbf{Variance Reduction}\\
\hline
$S_0=1$, $K=0.1$ & 1.09$\times 10^{11}$\\
$S_0=1$, $K=0.9$ & 3.70$\times 10^{11}$\\
$S_0=5$, $K=1$ & 4.37$\times 10^{11}$\\
$S_0=50$, $K=10$ & 4.37$\times 10^{11}$ \\
$S_0=50$, $K=49$ & 1.67$\times 10^{12}$
\end{tabular}
} \vspace{0.25cm}
\caption{Variance Reduction of an SMC Method against Importance Sampling. This is for an arithmetic Asian
option, under the BNS SV model. The algorithms were both run 50 times.}%
\label{tab:iscomp}%
\end{table}

\section{Summary}\label{sec:summary}

In this article we have provided a summary of SMC methods and shown the
potential benefits for their application in option pricing and sensitivity
estimation. Many of the examples presented in this paper have been based
upon translations of algorithms that have been used in other contexts such
as stochastic control (an exception is Section \ref{sec:pricingasians}). On the basis
of the work here, it is felt that the application of SMC methods focussed
upon particular problems, is likely to be highly beneficial. 

The reader of this article should be cautious. There is no claim
that SMC methods will always work well, and indeed be uniformly superior to other Monte
Carlo methods, or even deterministic approaches. In the process of working
on the article, many financial models were too complex to apply standard
SMC methods. What
is claimed, at least, is that the methods can help to push the boundaries
of models that may be used for pricing and is a useful technique for any
Monte Carlo specialist in option pricing.

\subsubsection*{Acknowledgements}
We would like to thank Dr.~M.~Gander, Maple-Leaf Capital London UK, who provided some useful insights into real-time option pricing. Also to Prof.~D.~Stephens
and Prof.~A.~Doucet
for some useful conversations related to this work. 

\section*{Appendix}

In this appendix the proof of Proposition \ref{prop:recursion} as well as
the model and MCMC method that was used in Section \ref{sec:pricingasians}
are  given. 

\subsection*{Proposition \ref{prop:recursion}}
\begin{proof}
A more general proof, by induction, is given. 
Let $m=2$, and $f\in\mathcal{B}_b(E_2)$
(the bounded measurable functions on $E_2$), then:
\begin{eqnarray*}
\widetilde{\Lambda}_2(f) & = & \int_{E_1} \frac{\partial}{\partial\theta}\{
p_{\theta}(x_1|x_0)\}
\widetilde{\Phi}(s_{t_1}) \bigg[\int_{E_2} f(x_2) p_{\theta}(x_2|x_1)
\widetilde{\Phi}(s_{t_2})dx_2\bigg]dx_1 + \\ & & 
\int_{E_1}p_{\theta}(x_1|x_0)\widetilde{\Phi}(s_{t_1})
\bigg[\int_{E_2} f(x_2)\frac{\partial}{\partial \theta}\{p_{\theta}(x_2|x_1)\}
\widetilde{\Phi}(s_{t_2})dx_2\bigg]dx_1
\end{eqnarray*}
application of the Fubini theorem gives
\begin{eqnarray*}
\widetilde{\Lambda}_2(f) & = & 
\int_{E_1\times E_2} 
\widetilde{\Phi}(s_{t_1})\widetilde{\Phi}(s_{t_2})f(x_2)
\frac{\partial}{\partial\theta}\{
p_{\theta}(x_1|x_0)\}p_{\theta}(x_2|x_1)
dx_{1:2}
+ \\ & &  \int_{E_1\times E_2} 
\widetilde{\Phi}(s_{t_1})\widetilde{\Phi}(s_{t_2})f(x_2)
p_{\theta}(x_1|x_0)\frac{\partial}{\partial\theta}\{p_{\theta}(x_2|x_1)\}
dx_{1:2}\\
& = & \int_{E_1\times E_2}\widetilde{\Phi}(s_{t_1})\widetilde{\Phi}(s_{t_2})f(x_2)
\frac{\partial}{\partial\theta}\{p_{\theta}(x_2|x_1)p_{\theta}(x_1|x_0)\}dx_{1:2}\\
& = & \Lambda_{\theta}(\Phi f,s_0).
\end{eqnarray*}
Now assume that the identity holds for $m=n$ and consider $n+1$, $f\in\mathcal{B}_b(E_{n+1})$:
\begin{eqnarray*}
\Lambda_{\theta}(\Phi f, s_0) & = & \int_{E_{[1,n+1]}} \prod_{i=1}^{n+1}\widetilde{\Phi}(s_{t_i})
f(x_{n+1})
\frac{\partial}{\partial \theta}\{\prod_{i=1}^{n+1}p_{\theta}(x_i|x_{i-1})\}dx_{1:n+1}\\
& = & \int_{E_{[1,n+1]}} \prod_{i=1}^{n+1}\widetilde{\Phi}(s_{t_i})
f(x_{n+1})\frac{\partial}{\partial \theta}\{p_{\theta}(x_{n+1}|x_{n})\}
\prod_{i=1}^{n}p_{\theta}(x_i|x_{i-1})\}dx_{1:n+1} + \\ & & 
\int_{E_{[1,n+1]}} \prod_{i=1}^{n+1}\widetilde{\Phi}(s_{t_i})
f(x_{n+1})p_{\theta}(x_{n+1}|x_{n})
\frac{\partial}{\partial\theta}\bigg\{\prod_{i=1}^{n}p_{\theta}(x_i|x_{i-1})
\bigg\}dx_{1:n+1}\\
& = & \int_{E_{[n,n+1]}}\widetilde{\Phi}(s_{t_{n+1}}) f(x_{n+1})
\frac{\partial}{\partial \theta}\{p_{\theta}(x_{n+1}|x_{n})\}
\bigg[\int_{E_{[1,n-1]}}\bigg\{\prod_{i=1}^{n}p_{\theta}(x_i|x_{i-1})
\times \\ & & 
\widetilde{\Phi}(s_{t_i})
\bigg\}dx_{1:n-1}\bigg]
dx_{n:n+1}+ 
\int_{E_{[n,n+1]}}\widetilde{\Phi}(s_{t_{n+1}}) f(x_{n+1})
p_{\theta}(x_{n+1}|x_{n})\\ & &
\bigg[\int_{E_{[1,n-1]}}\bigg\{\prod_{i=1}^{n}\widetilde{\Phi}(s_{t_i})\bigg\}
\frac{\partial}{\partial\theta}\bigg\{\prod_{i=1}^{n}p_{\theta}(x_i|x_{i-1})
\bigg\}dx_{1:n-1}\bigg]dx_{n:n+1}\\
& = & \int_{E_{[n,n+1]}}\widetilde{\Phi}(s_{t_{n+1}}) f(x_{n+1})
\frac{\partial}{\partial \theta}\{p_{\theta}(x_{n+1}|x_{n})\}\Pi_n(dx_n)dx_{n+1}
+ \\ & & \int_{E_{[n,n+1]}}\widetilde{\Phi}(s_{t_{n+1}}) f(x_{n+1})
p_{\theta}(x_{n+1}|x_{n})\widetilde{\Lambda}_n(dx_n)dx_{n+1}\\
& = & \widetilde{\Lambda}_{n+1}(f)
\end{eqnarray*}
setting $f\equiv 1$ completes the proof. 
\end{proof}

\subsection*{Model}
Consider a finite collection of time-points $0=t_0<t_1<\cdots<t_m=T$, then the
joint density of the log-price and integrated volatilities are:
$$
\bigg\{\prod_{i=1}^m \phi\{y_{t_i};y_{t_{i-1}} + \mu(t_{i}-t_{i-1}), \tilde{\sigma}_i(v_{t_{1}:t_i})\}
p(v_{t_i})\bigg\}
$$
where $\phi(\cdot;\mu,\sigma)$ is the normal density of mean $\mu$ and variance
$\sigma$ and, with $i\in\mathbb{T}_m$, $v_{t_i}=(a_{1:n_i}^i,r_{1:n_i}^i,n_i)\in
[0,\lambda\nu(t_i-t_{i-1})]^{n_i}\times[0,1]^{n_i}\times\mathbb{N}$ is a 2-dimensional Poisson process on $[0,\lambda\nu(t_{i}-t_{i-1})]\times [0,1]$
of unit rate. The functions $\tilde{\sigma}_i(\cdot)$ are as follows.
Let $\bar{\sigma}_0=v_0$ and define
\begin{eqnarray*}
\gamma_{i,1} & := & e^{-\lambda(t_i-t_{i-1})}\sum_{j=1}^{n_i}\log\bigg(\frac{\lambda\nu(t_i-t_{i-1})}
{a_j^i}\bigg)e^{\lambda(t_i-t_{i-1})r_{j}^i}\\
\gamma_{i,2} & := & \sum_{j=1}^{n_i}\log\bigg(\frac{\lambda\nu(t_i-t_{i-1})}
{a_j^i}\bigg)
\end{eqnarray*}
then
\begin{eqnarray*}
\bar{\sigma}_i(v_{t_1:t_i}) & = & e^{-\lambda(t_i-t_{i-1})}\bar{\sigma}_{i-1}(v_{t_1:t_i}) + \gamma_{i,1}
\\
\tilde{\sigma}_i(v_{t_1:t_i}) & = & \gamma_{i,2} - \bar{\sigma}_i(v_{t_1:t_i}) + \bar{\sigma}_{i-1}(v_{t_1:t_i}).
\end{eqnarray*}

\subsection*{MCMC}
The moves are based upon Metropolis-Hastings kernels; see \cite{robert} for
an introduction.
There are two main quantities to be updated; the $\nu_{t_{1}:t_{m}}$ and the parameters 
$$
(a_{1:n_1}^1,r_{1:n_1}^1,n_1,\dots, a_{1:n_m}^m,r_{1:n_m}^m,n_m)
$$ 
that make up the volatility process. From here-in simplify the notation to $\nu_{1:m}$.

The  $\nu_{1:m}$ are updated, by picking an $i\in\mathbb{T}_m$, uniformly at random, and sampling from its process density. The move is accepted or rejected, according to a standard Metropolis-Hastings acceptance probability:
$$
1\wedge \frac{\varphi_{\nu_i'}\{\nu_{i+1};\tilde{\mu}(\nu_{i-1},\nu_{i}'),\tilde{\sigma}_{i+1}(v_{1:i+1})\}}
{\varphi_{\nu_i}\{\nu_{i+1};\tilde{\mu}(\nu_{i-1:i}),\tilde{\sigma}_{i+1}(v_{1:i+1})\}}
\times
\frac{\varphi_{\nu_{i+1}}\{\nu_{i+2};\tilde{\mu}(\nu_{i}',\nu_{i+1}),\tilde{\sigma}_{i+2}(v_{1:i+2})\}}
{\varphi_{\nu_{i+1}}\{\nu_{i+2};\tilde{\mu}(\nu_{i:i+1}),\tilde{\sigma}_{i+2}(v_{1:i+2})\}}
$$
with $1\leq i \leq m-2$; a similar formula can be calculated if $i\in\{m-1,m\}$.

The $(a_{1:n_1}^1,r_{1:n_1}^1,n_1,\dots, a_{1:n_m}^m,r_{1:n_m}^m,n_m)$ are
updated in a similar way to \cite{griffin}. Again, pick an $i\in\mathbb{T}_m$ uniformly at random. Then, the number of points, $n_i$ is either
increased by 1 (birth), or decreased, if possible, by 1 (death); the choice is made at random.
If a birth occurs, the new $a_i,r_i$ are sampled according to the process
density. The Metropolis-Hastings acceptance probability, for a birth is
$$
1 \wedge \frac{\prod_{j=i}^m \varphi_{\nu_{j-1}}\{\nu_{j};\tilde{\mu}(\nu_{j-2:j-1}),
\tilde{\sigma}_j(v_{1:i},,v_{i:j}')\}}
{\prod_{j=i}^m \varphi_{\nu_{j-1}}\{\nu_{j};\tilde{\mu}(\nu_{j-2:j-1}),
\tilde{\sigma}_j(v_{1:j})\}}\times \frac{\lambda\nu(t_{i}-t_{i-1})d(n_{i}+1)}{b(n_i)}
$$
where $b(\cdot), d(\cdot)$ are the probabilities of proposing
birth and death moves. The death move, when $n_i=n_i+1$, has  a ratio that
is the inverse of that above.

\end{document}